\documentclass[aps,prx,preprint,superscriptaddress]{revtex4-1}

\usepackage[version=3]{mhchem}
\usepackage{bm}
\usepackage{amsmath}
\usepackage{ulem}
\usepackage{amssymb}
\usepackage{amsfonts}
\usepackage{euscript}
\usepackage{color}
\usepackage{epsfig}
\usepackage[section]{placeins}
\usepackage{gensymb}
\usepackage{array}
\usepackage{mathtools}

\makeatletter
\newcommand{\thickhline}{
    \noalign {\ifnum 0=`}\fi \hrule height 1pt
    \futurelet \reserved@a \@xhline
}
\newcolumntype{"}{@{\hskip\tabcolsep\vrule width 1pt\hskip\tabcolsep}}
\makeatother

\begin{document}

\title{Hybridization and the origin of Fano resonances  in\\symmetric nanoparticle trimers}

\author{Ben~Hopkins}
\email{ben.hopkins@anu.edu.au}
\affiliation{Nonlinear Physics Centre, Australian National University, Acton, ACT 2601, Australia}

\author{Dmitry~S.~Filonov}
\affiliation{ITMO University, St. Petersburg 197101, Russia}

\author{Stanislav~B.~Glybovski}
\affiliation{ITMO University, St. Petersburg 197101, Russia}

\author{Andrey~E.~Miroshnichenko}
\email{andrey.miroshnichenko@anu.edu.au}
\affiliation{Nonlinear Physics Centre, Australian National University, Acton, ACT 2601, Australia}


\begin{abstract}
We study the light scattering by plasmonic and dielectric symmetric trimers to investigate the existence of polarization-independent Fano resonances.
Plasmonic hybridization theory is revealed to hide simple physics, and we instead provide a simplified model for hybridization to derive a plasmonic trimer's eigenmodes analytically.  
This approach is demonstrated to accurately recreate full wave simulations of plasmonic trimers and their Fano resonances.
We are subsequently able to deduce the grounds for modal interference in plasmonic trimers and the related formation of Fano resonances.
However, by taking advantage of the generality of our simplified hybridization approach, we are also able to investigate the eigenmodes of all-dielectric trimers.
We show that bianisotropic coupling channels between high-index dielectric nanoparticles are able to increase the capacity for Fano resonances, even at normal incidence.   
We finally provide the first experimental measurements of sharp, polarization-independent Fano resonances from a symmetric all-dielectric trimer, with very good agreement to the predicted response from our simplified hybridization theory.  
\end{abstract}

\maketitle

\section{Introduction}

The Fano resonance~\cite{Miroshnichenko2010} has become a well-recognized interference feature in the optical scattering response of many nanoscale structures. 
It attracted considerable attention due to its distinctively-sharp extinction lineshape, but also due to anomalous near-field behavior for surface-enhanced Rahman scattering~\cite{Ye2012}, nonlinear response~\cite{Zhang2013}, and enhancement of circular dichroism~\cite{Hopkins2014}.
For plasmonic structures, the existence of optical Fano resonances has generally been explained through the use of a hybridization theory argument~\cite{Prodan2003}; an argument which involves subdividing a structure into two or more subsystems with known properties, and then deducing how their optical responses combine together. 
The way such optical responses combine, or hybridize, is regularly depicted as per molecular orbital theory: the modes of each subsystem are added constructively or destructively to form a bonding and an antibonding mode, the latter of which exhibits suppressed scattering associated with the Fano resonance.  
For the case of plasmonics, this involves treating individual nanoparticles as electron density distributions; a model which inherently accounts for quite comprehensive physics and subsequently requires nontrivial derivations for even the most simple systems, such as concentric spheres~\cite{Prodan2004} and two-particle dimers~\cite{Nordlander2004}.
Using this model, even the comparatively simple case for plasmonic hybridization of a symmetric three-particle trimer has yet to be found.   
A consequence of this complexity is that hybridization is rarely performed explicitly.  
It has instead become a conceptual explanation for experimental and numerical observations in even quite complicated scattering systems for which hybridization solutions have never been found.     
Here we instead begin from the premise that the dominant optical properties of nanoparticle geometries do not require the level of complexity found in plasmonic hybridization theory.  
We propose a simplified approach to hybridization when modeling a nanoparticle system in the dipole approximation, thereby considering only the dipole responses of individual nanoparticles and the coupling between them.  
To demonstrate the capacity of this simplification, we approach the outstanding problem of modal hybridization in symmetric trimers.
In particular, the optical feature we focus on in symmetric trimers is the polarization-independent Fano resonance.  
This type of feature has previously only been observed in more complicated nanoparticle oligomers~\cite{Hentschel2010} where a conceptual hybridization can be performed in equivalent ways for arbitrary polarization angles.  
While the polarization independence of symmetric trimers is known~\cite{Brandl2006, Chuntonov2011, HopkinsLiu2013, Rahmani2013}, the existence of Fano resonances~\cite{Miroshnichenko2012, HopkinsPoddubny2013} and the indeed their combination with polarization indepedence is not obvious in terms of hybridization, particularly because it is not possible to subdivide a trimer in a way that conserves its symmetry.
Here we are able to resolve this problem by performing the simplified hybridization procedure to find the eigenmodes of a plasmonic trimer, analytically.
We show that the doubly-degenerate eigenmodes of trimers, those responsible for polarization indepedent scattering, become inherently nonorthogonal in the presence of retarded coupling between particles.  
Such eigenmodes can therefore allow polarization-independent Fano resonances at normal incidence~\cite{HopkinsPoddubny2013}, and we provide full wave simulations to demonstrate these Fano resonances in plasmonic nanoparticle trimers, with good agreement to our analytical eigenmode results.  
However, by recognizing that our simplified approach to hybridization is portable to other models, we further investigate all-dielectric trimers in the coupled electric and magnetic dipole approximation~\cite{Mulholland1994} to account for the existence of magnetic dipolar responses of individual high-index dielectric nanoparticles. 
This analysis shows that previously reported bianisotropic coupling effects in dielectric oligomers~\cite{HopkinsFilonov2015} also play an important role in the optical response of all-dielectric geometries at normal incidence.  
The result is then a three-fold increase in the number of doubly-degenerate eigenmodes that can be excited by a normally-incident plane wave in an all-dielectric trimer, when compared to the plasmonic trimer.   
Moreover, all these eigenmodes are nonorthogonal and subsequently lead to a substantial increase in the presence and magnitude of polarization-independent Fano resonances at normal incidence. 
We then verify these predictions in a radio frequency experiment that mimics the optical transmission through a silicon nanosphere trimer.
We observe a number of sharp, polarization-independent Fano resonances from the all-dielectric trimer at normal incidence. 
The eigenmodes calculated from our simplified hybridization approach are then able to provide a quantitative recreation of these experimental measurements and their explicit eigenmode decompositions.
In doing so, we can unambiguously demonstrate the validity of our simplified approach to hybridization.

\section{The eigenmodes of plasmonic trimers}

\begin{figure*}[!th]
\centering
\centerline{\includegraphics[width=0.95\textwidth]{{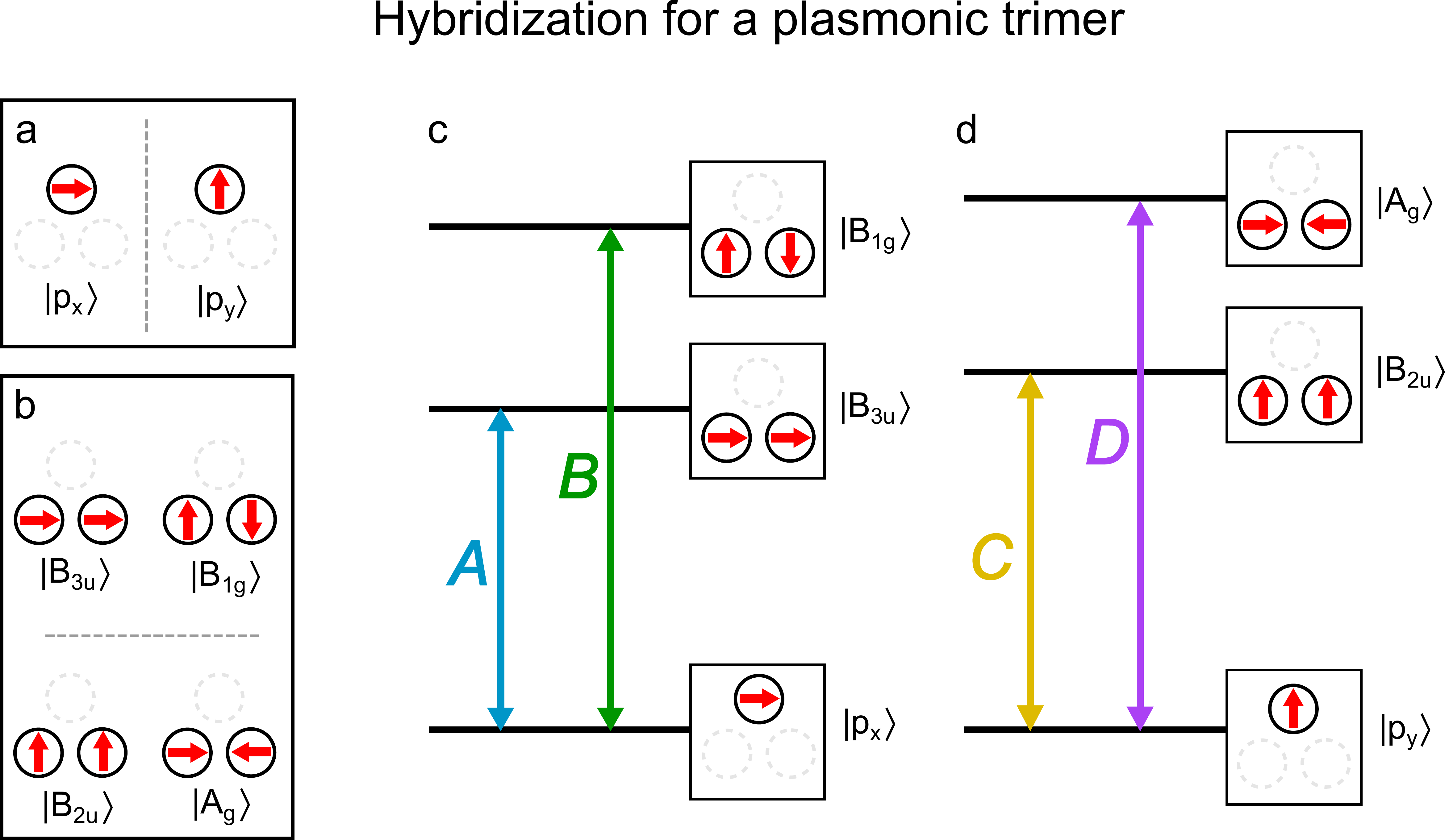}}}
\caption{The eigenmodes of (a) a single particle and (b) a dimer with {$\mathrm{D_{2h}}$} symmetry, when assuming only in-plane electric dipole responses from each particle.  The dimer eigenmodes are labeled according to their associated irreducible representation. We also show the location of the extra particle/s needed to form a trimer. In (c) and (d), we depict a generalized energy level diagram for these eigenmodes and identify the coupling channels, labeled as $A$-$D$. }
\label{fig:electric_eigenmodes}
\end{figure*}

For the purpose of performing hybridization, a trimer can only be subdivided into a dimer and a single particle; the other option would be the trivial case of dividing it into three single particles.  
Here, we restrict ourselves to considering only the electric dipole response of each individual particle, since it  dominates for subwavelength plasmonic particles.  
We can therefore utilize the dipole approximation~\cite{Draine1994}, where each particle's dipole moment $(\mathbf{p}_{i})$ is related to the externally-applied electric field $(\mathbf{E_0})$ as:
\begin{align}
\mathbf{p}_{i}  = \alpha_{\scriptscriptstyle E}\epsilon_{0}\mathbf{E_{0}}(\mathbf{r}_{i})&+\alpha_{\scriptscriptstyle E}k^{2}
\sum \limits_{j\neq i} \hat{G}_{0}(\mathbf{r}_{i},\mathbf{r}_{j}) \cdot \mathbf{p}_{j}
\label{eq:dipole equation}
\end{align}
Here, $\alpha_{\scriptscriptstyle E}$ is the electric dipole polarizability of a particle and $\hat{G}_{0}$ is the free space dyadic Green's function, which acts on dipole moments as:
\begin{align}
\hat{G}_0(\mathbf{r'},\mathbf{r})\cdot \mathbf{p} 
&= \frac{e^{i k R }}{4 \pi   R}\left[ \left(1 + \frac{i}{kR}- \frac{1}{k^2 R^2}\right) \mathbf{p} - \left( 1 + \frac{3i}{kR}- \frac{3}{k^2R^2}\right) (\mathbf{n}\cdot\mathbf{p})\mathbf{n}  \right ] \nonumber
\end{align}
where $k$ is the wavenumber, $\mathbf{n}$ is the unit vector pointing from $\mathbf{r}$ to $\mathbf{r'}$ and $R = |\mathbf{r}-\mathbf{r'}|$.
In this model, the single particle and dimer eigenmodes are both known and easily deduced, and are shown in Fig.~\ref{fig:electric_eigenmodes}a and \ref{fig:electric_eigenmodes}b.
To derive how the dimer and single particle hybridize, we have to calculate the coupling channels between each dimer and single particle eigenmode, labeled  as $A$-$D$ in Fig.~\ref{fig:electric_eigenmodes}c and \ref{fig:electric_eigenmodes}d.  
To find the coupling channels, we use Eq.~\ref{eq:dipole equation} to calculate the dipole moments induced in the dimer by a single particle eigenmode and project those onto the dimer eigenmodes (which form a complete basis for the dimer's response), and {\it vice versa}.  
This gives us expressions for the coupling channels shown in Fig.~\ref{fig:electric_eigenmodes}c.
\begin{subequations}
\begin{align}
A  &\;=\;   \frac{\alpha_{\scriptscriptstyle E} e^{i k R }}{8  \sqrt{2}\,\pi  R}\left[ 3 k^{2} + \frac{ik}{R}- \frac{1}{R^2}\right ]    \\
B  &\;=\;  - \frac{\sqrt{3}\,\alpha_{\scriptscriptstyle E}  e^{i k R }}{8 \sqrt{2}\, \pi  R}\left[k^{2}+ \frac{3ik}{R}- \frac{3}{R^2}\right ] \\
C  &\;=\;  \frac{\alpha_{\scriptscriptstyle E} e^{i k R }}{8 \sqrt{2}\,\pi  R}\left(  k^{2} -  \frac{5 ik}{R} + \frac{5}{R^2}\right )   \\
 D  &\;= \; B
\end{align}
\label{eq:coupling}
\end{subequations}
where we have normalized the single particle and dimer modes to make all channels omnidirectional.  
All other coupling channels between the eigenmodes of the dimer and those of the single particle are zero, which is due to a geometry-induced symmetry mismatch between basis vectors.  
It is important to acknowledge that, unlike molecular orbital hybridization, the hybridization of scattering structures does not have a perennial set of basis vectors ({\it i.e.} atomic orbitals) with known symmetry mismatch conditions to identify recurring coupling/hybridization channels in different geometries.  
In nanostructures, the coupling channels have to be derived for each given geometry in order to perform hybridization correctly.     
Importantly, our formulation of hybridization also relies on the basis vectors being eigenmodes of their associated subsystem.
Indeed, if this was {\it not} the case, we would have to account for new coupling channels arising from interactions within each subsystem, which would substantially increase the number of coupling channels we have to consider and also their complexity.   
In any case, to now find the trimer's hybridized eigenmodes we recognize that an eigenmode's dipole moment profile must satisfy the dipole equation (Eq.~\ref{eq:dipole equation}) as:
\begin{align}
\mathbf{v}_i = \alpha_{\scriptscriptstyle E} \epsilon_0 \lambda \mathbf{v}_i  +  \sum \limits_{j\neq i} \alpha_{\scriptscriptstyle E}k^2 \hat{G}_0(\mathbf{r}_i,\mathbf{r}_j)\cdot \mathbf{v}_j
\label{eq:eig equation}
\end{align}
We can then rewrite Eq.~\ref{eq:eig equation} as a system of linear equations by considering the dimer and single particle eigenmodes ({\it i.e.} in  Fig.~\ref{fig:electric_eigenmodes}) as orthogonal basis vectors.  
Moreover, we can consider an eigenmode, $\left| {v_i} \right \rangle$, as a linear combination of these basis vectors:
\begin{align}
\left| {v_i} \right \rangle= a_i \left|\mathrm{p_x}\right \rangle + b_i\left|  \mathrm{B_{3u}}\right \rangle  + c_i \left|\mathrm{B_{1g}}\right \rangle +  d_i \left|\mathrm{p_y}\right \rangle + e_i\left|  \mathrm{B_{2u}}\right \rangle  + f_i \left|\mathrm{A_{g}}\right \rangle 
\label{eq:linear combination}
\end{align}
As detailed in Appendix~\ref{appendixA}, this basis allows us to solve Eq.~\ref{eq:eig equation} as a matrix equation to find analytical expressions for the hybridized eigenmodes of the trimer.  
Specifically, we obtain the following expressions for the hybridized eigenmodes:
\begin{subequations}\label{eq:result}
\begin{align}
\left| {v_1} \right \rangle:& \left \{ 
\begin{array}{ll}
a_1 &=\sqrt{2} \\
b_1 &= -1 \\
c_1 &= \sqrt{3}\\
 \lambda_1 &=  (\alpha_{\scriptscriptstyle E} \epsilon_0)^{-1} + \frac{e^{i k R }}{8 \pi \epsilon_0  R} \left(3 k^{2} + \frac{5ik}{R}- \frac{5}{R^2}\right)
\end{array}\right.
\\[2ex] 
\left| {v_{2x}} \right \rangle,\left| {v_{3x}} \right \rangle:& \left \{
\begin{array}{ll}
a_2, a_3 &= \frac{\sqrt{3}e^{i k R }}{4 \sqrt{2} \pi \epsilon_0  R} \left(5 k^{2} + \frac{3ik}{R}- \frac{3}{R^2} \right) 
\pm 2\sqrt{6}\;\delta 
 \\[1ex]
b_2, b_3 &=  \frac{\sqrt{3}e^{i k R }}{2  \pi \epsilon_0  R} \left( k^{2} - \frac{3ik}{R}+ \frac{3}{R^2} \right) 
\pm  4\sqrt{3}\;\delta 
\\[1ex]
c_2, c_3 &= -\frac{3e^{i k R }}{4  \pi \epsilon_0  R} \left( k^{2} + \frac{3ik}{R} - \frac{3}{R^2} \right) 
\\[1ex]
\lambda_{2},\lambda_{3} &= \lambda_0 \mp  \delta
\end{array}\right.\\[2ex]
\left| {v_{2y}} \right \rangle,\left| {v_{3y}} \right \rangle:& \left \{
\begin{array}{ll}
d_2, d_3 &=\frac{\sqrt{3}e^{i k R }}{4 \sqrt{2} \pi \epsilon_0  R} \left( k^{2} - \frac{9ik}{R}+ \frac{9}{R^2} \right) 
 \pm  2\sqrt{6}\;\delta 
 \\ [1ex]
e_2, e_3 &=\frac{\sqrt{3} k^{2} e^{i k R }}{ \pi \epsilon_0  R} 
\pm  4\sqrt{3}\;\delta 
 \\[1ex]
f_2, f_3 &=  -\frac{3e^{i k R }}{4  \pi \epsilon_0  R} \left( k^{2} + \frac{3ik}{R} - \frac{3}{R^2} \right) 
\\[1ex]
\lambda_{2},\lambda_{3} &=\lambda_0 \mp \delta
\end{array}\right.\\[2ex]
\left| {v_4} \right \rangle:& \left \{ 
\begin{array}{ll}
d_4 &=\sqrt{2} \\
e_4 &= -1 \\
f_4 &= -\sqrt{3}\\
 \lambda_4 &=  (\alpha_{\scriptscriptstyle E} \epsilon_0)^{-1} - \frac{e^{i k R }}{8 \pi \epsilon_0  R} \left(  k^{2} + \frac{7ik}{R}- \frac{7}{R^2}\right)
\end{array}\right. \label{eq:v4}
\end{align}
\label{eq:plasmonic eigenmodes}
\end{subequations}
Here all undefined coefficients ({\it cf.} Eq.~\ref{eq:linear combination}) are zero, and  we have defined a central eigenvalue, $\lambda_0$, for  the $\left| {v_{2}} \right \rangle$ and $\left| {v_{3}} \right \rangle$ eigenmodes, along with a splitting function, $\delta$, that produces their nondegeneracy.
\begin{align}
\lambda_0 &\;\coloneqq\;   (\alpha_{\scriptscriptstyle E} \epsilon_0)^{-1} - \frac{e^{i k R }}{16 \pi \epsilon_0  R} \left(k^2 - \frac{ik}{R}+ \frac{1}{R^2}\right) \\
\delta &\;\coloneqq\;  \frac{1 }{\alpha_{\scriptscriptstyle E} 16 \pi \epsilon_0  R}\sqrt{5 \alpha_{\scriptscriptstyle E} ^{2}e^{2 i k R }\left[\left( \frac{k^{2}}{5} + \frac{3ik}{R}- \frac{3}{R^2} \right)^{2}+\left( \frac{8 k^2}{5}\right)^{2}\right]} \label{eq:delta}
\end{align}
We can categorize the hybridized eigenmodes according to their irreducible representation in the trimer's $\mathrm{D_{3h}}$ point group (see Table~\ref{tab:D3h}) .
The $\left| {v_1} \right \rangle$ eigenmode has azimuthally-oriented dipole moments and the $\left| {v_4} \right \rangle$ eigenmode has radially-oriented dipole moments, making them the trimer's $\mathrm{A'_2}$ and $\mathrm{A'_1}$ eigenmodes, respectively~\cite{Brandl2006}.  On the other hand, the $\left| {v_{2}} \right \rangle$  and $\left| {v_{3}} \right \rangle$ eigenmodes are the, doubly-degenerate, $\mathrm{E'}$ eigenmodes, which are responsible for polarization-independent behavior.
Because we have obtained two distinct $\mathrm{E'}$ eigenmodes, which  is known to be the maximum number that a ring-type oligomer can exhibit in the dipole approximation~\cite{HopkinsPoddubny2013}, we know that the hybridization approach here has determined {\it all} of the $\mathrm{E'}$ doubly-degenerate eigenmodes for a plasmonic trimer.  
\begin{table}
\begin{tabular}{c | c  c  c  c  c  c  c} 
~ & $\,E\,$  & $\,2 C_{3}\,$    & $\,3 C'_2\,$  &  $\,\sigma_h\,$  & $\,2 S_3\,$ & $\,3 \sigma_v \,$ \\  
\hline \\[-4ex]
$\mathrm{A'_{1}}\;$ & 1 &  1 &  1&  1  &  1 &  1  \\
$\mathrm{A'_{2}}\;$ & 1 &  1 &  -1& 1  &  1 &  -1  \\
$\mathrm{E'}\;$        & 2 &  -1 & 0&  2 &  -1 &  0    \\
$\mathrm{A''_{1}}\;$ & 1 &  1 &  1&  -1 & -1 & -1   \\
$\mathrm{A''_{2}}\;$ & 1 &  1 & -1& -1 & -1 & 1\\
$\mathrm{E''}\;$         & 2 & -1 & 0& -2 &  1 & 0   \\
 [0.5ex] 
\hline 
\end{tabular}\\~
\caption{Character table for the $\mathrm{D_{3h}}$ symmetry group. The rows correspond to different irreducible representations and the columns are the symmetry operations.  Each number is the trace of the associated operation's matrix representation.~\cite{Dresselhaus2008}}
\label{tab:D3h}
\end{table}

\begin{figure}
\centering
\centerline{\includegraphics[width=0.85\textwidth]{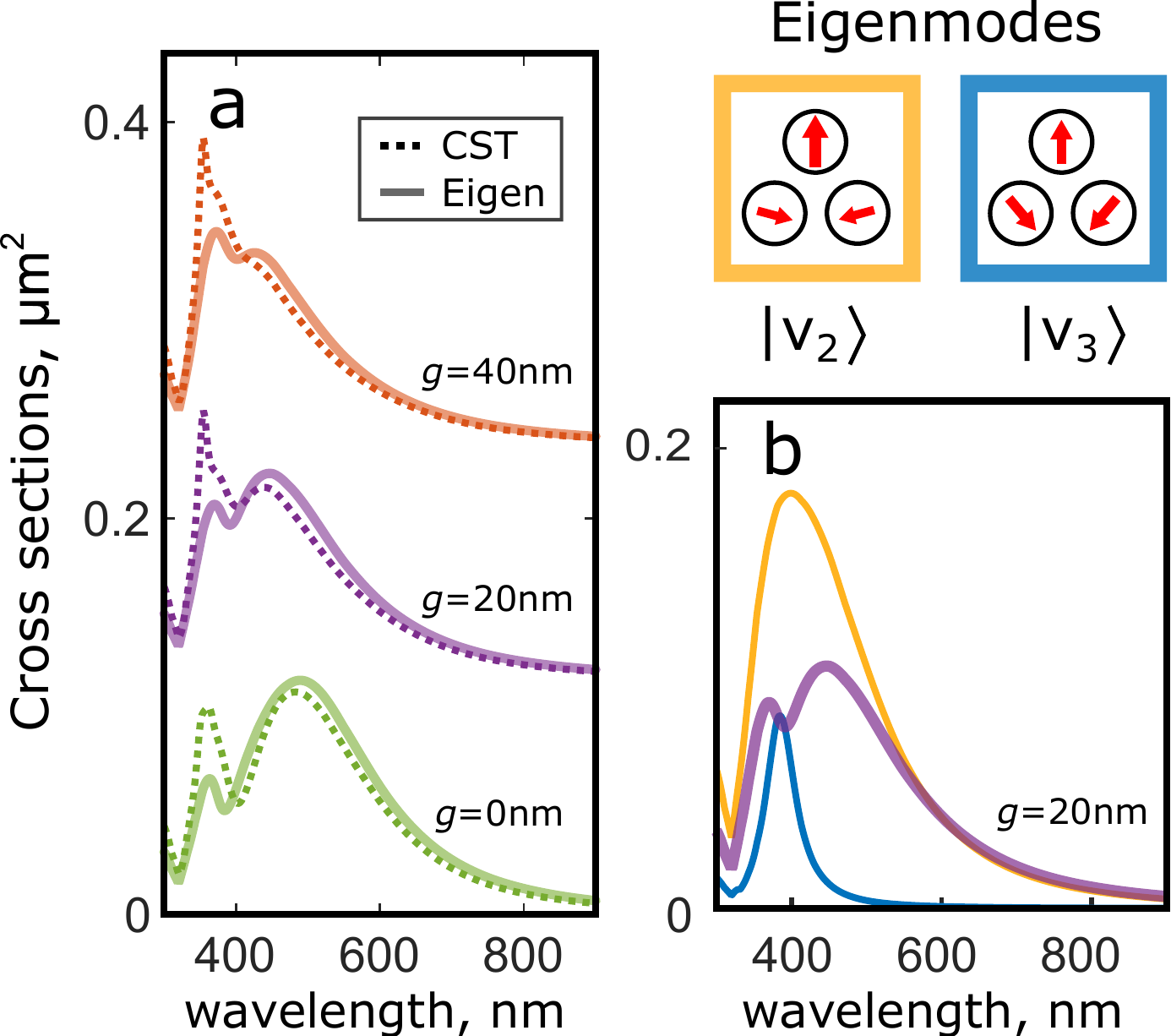}}
\caption{(a) Extinction cross section from a trimer made of 100nm silver nanospheres, simulated using (dashed line) CST Microwave Studio and (unbroken line) the analytical eigenmodes of Eq.~\ref{eq:plasmonic eigenmodes}.  Both curves show the formation of a small Fano resonance as the gap between particles, {\it g}, is reduced.  (b) Decomposition of the extinction for the {{\it g}$\,=\,$20nm} trimer in terms of the two excited eigenmodes.  The eigenmode profiles depict the real components of dipole moments at 400nm.}
\label{fig:plasmonictrimer}
\end{figure}

It is worth pausing to acknowledge an interesting result of this analysis.  
If we were to consider the doubly-degenerate eigenmodes in complex frequency space ($s$-space) to find and/or analyze their resonances~\cite{Powell2014}, there are points in $s$-space where the expressions for $\left| {v_{2}} \right \rangle$ and $\left| {v_{3}} \right \rangle$ are the same, specifically: when $\delta=0$.  
At such points, the two eigenmodes coalesce and the eigenspace of these two eigenmodes subsequently reduces in dimension, indicating a nontrivial topology of the eigenspace.    
Such coalescence points are also known as {\it exceptional points} and the coalescing eigenmodes are known to interchange expressions with each other, depending on the path taken through $s$-space in the vicinity of these points~\cite{Heiss2001, Dembowski2001, Heiss2012}. 
In a trimer, the locations of four such exceptional points can be found by solving $\delta=0$ for a complex $k$ in Eq.~\ref{eq:delta}.
An investigation into eigenspace topology is well beyond the scope of this work and, instead, it suffices that the eigenmodes and eigenvalues in Eq.~\ref{eq:result} are {\it exact} expressions for that of a symmetric trimer when considering only the electric dipole response of each individual nanoparticle.
Furthermore, using the expressions in Eq.~\ref{eq:result}, it follows that, in general, $\left| {v_{2}} \right \rangle$ and $\left| {v_{3}} \right \rangle$ will be nonorthogonal and written in terms of the eigenmode's basis vector coefficients:
\begin{align}
\langle v_{2}|v_{3}\rangle \,=\; a_2^* a_3 + b_2^*b_3 +c_2^*c_3 \;=\;  d_2^* d_3 + e_2^*e_3 +f_2^*f_3
\label{eq:plasmonic overlap}
\end{align}
This means that polarization-independent Fano resonances could indeed be expected~\cite{HopkinsPoddubny2013} to occur in plasmonic trimers, which is interesting given neither the dimer nor single particle are able to exhibit Fano resonances in isolation.  
In other words, the presence of coupling channels between dimer and single particle have directly led to nonorthogonality in the trimer's hybridized eigenmodes.
This can be understood conceptually as the single particle providing a coupling channel into the $|\mathrm{A_{g}}\rangle$ and $|\mathrm{B_{1g}}\rangle$ basis vectors, which are orthogonal to the incident field and thereby act as a damped oscillators that are coupled with the remainder of the system.  
However, as is discussed in Appendix~\ref{appendixA}, it is also important that the coupling channels provide retardation in propagation for Eq.~\ref{eq:plasmonic overlap} to not sum to zero.
In any case, to now demonstrate Fano resonances of the nature we are describing, we can consider a symmetric trimer made of silver nanoparticles with a diameter of 100~nm, and we then vary the gap between particles from 40~nm down to touching. 
In Figure~\ref{fig:plasmonictrimer}, we show the evolution of the extinction cross section from this trimer for a normally-incident plane wave.
We provide two calculations for extinction: (dashed line) directly from CST Microwave Studio and (unbroken line) from the analytical expressions for the eigenmodes and eigenvalues in Eq.~\ref{eq:plasmonic eigenmodes}.
Notably, with the exception of the quadrupole resonance at high frequencies, there is a good match between the eigenmode analysis and the full wave simulations.  
Importantly, both approaches observe a small Fano resonance forming as the gap between particles is reduced.   
However, the advantage of using eigenmodes is we can attribute portions of the extinction to individual eigenmodes.
Indeed, in Fig.~\ref{fig:plasmonictrimer}b, we can see that the Fano resonance is produced by interference~\cite{HopkinsPoddubny2013} between the two eigenmodes.
Admittedly, while our model does not observe the single particle quadrupole response, an extension our hybridization approach to models that account for quadrupole responses, such as in Ref.~\cite{Evlyukhin2011}, is not unfeasible.
In any case, the results in Fig.~\ref{fig:plasmonictrimer} are able to validate the predictions and analysis of our approach to hybridization for a plasmonic trimer.

\section{The eigenmodes of all-dielectric trimers}

\begin{figure*}[!th]
\centering
\centerline{\includegraphics[width=0.95\textwidth]{{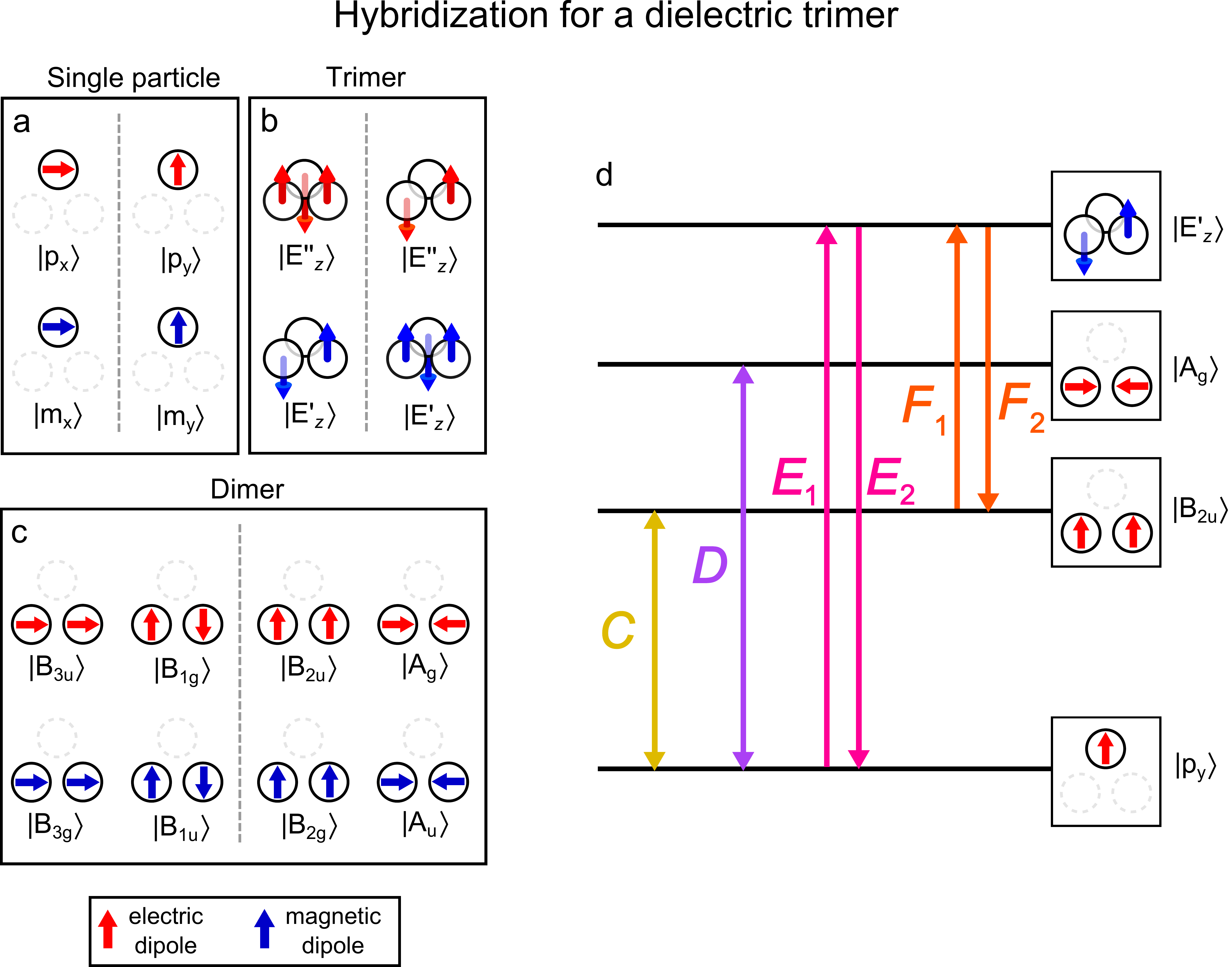}}}
\caption{The eigenmodes of (a) a single particle and (c) a dimer with $\mathrm{D_{2h}}$ symmetry, when assuming both electric and magnetic dipole responses from each particle, and neglecting the $z$-direction.  
The (b) $z$-oriented basis vectors of a trimer, which  become  doubly-degenerate eigenmodes of the trimer when electric-magnetic coupling is neglected.  
The dimer and trimer eigenmodes are labelled according to their associated irreducible representation. In (d) we depict a generalized energy level diagram analogous to that in Fig.~\ref{fig:electric_eigenmodes}d, now with coupling channels $C$-$F$   (see  Eqs.~\ref{eq:coupling} and \ref{eq:coupling' extra}) }
\label{fig:new_eigenmodes}
\end{figure*}

\begin{figure*}[!th]
\centering
\centerline{\includegraphics[width=\textwidth]{{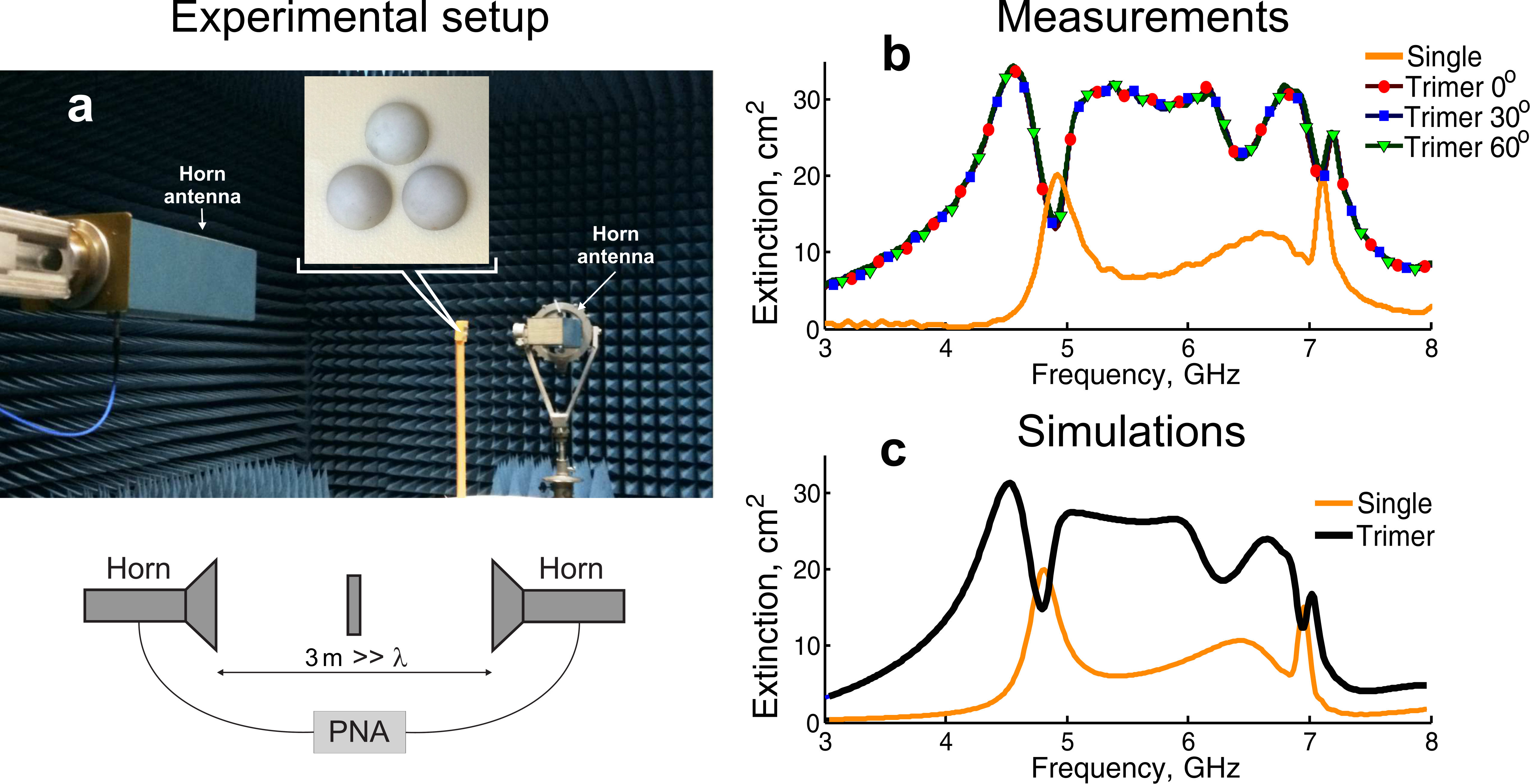}}}
\caption{ (a) Experimental setup for the trimer made from MgO-$\mathrm{TiO_2}$ spheres and (b) the  measurements of extinction for both the trimer and a single sphere. The trimer exhibits a pronounced, polarization-independent, Fano resonance at $4.8\,\mathrm{GHz}$ in very good agreement with  (c) the  simulation results of extinction, which were calculated using CST Microwave Studio.}
\label{fig:experiment}
\end{figure*}

\begin{figure}[!h]
\centering
\centerline{\includegraphics[width=0.6\columnwidth]{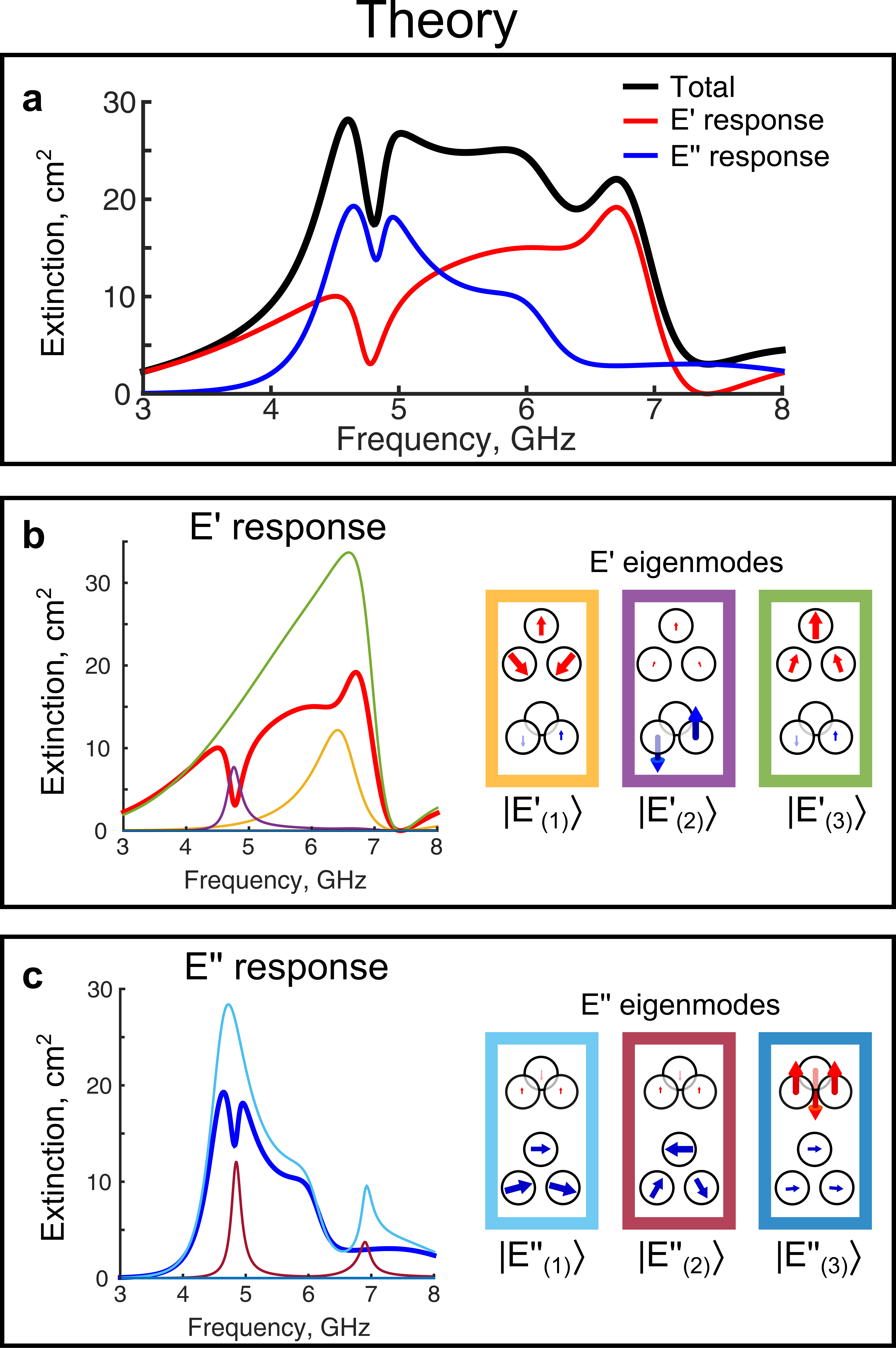}}
\caption{(a) Extinction of the trimer in Fig.~\ref{fig:experiment}, when calculated using the hybridization procedure presented in this paper.  We also show the eigenmode decomposition of the extinction coming from the $\mathrm{E'}$ and $\mathrm{E''}$ response of the trimer in (b) and (c).  Each eigenmode's isolated contribution to the extinction is shown alongside their associated dipole moment profiles, which depict the real components of each dipole moment at $4.8\,\mathrm{GHz}$.  Red arrows are electric dipole moments and blue arrows are magnetic dipole moments.}
\label{fig:theory}
\end{figure}

The analysis of the previous section corresponded to plasmonic nanoparticle trimers because we assumed only electric dipolar responses from individual nanoparticles.
However, strong Fano resonances were predicted to exist in silicon, {\it all-dielectric}, trimers~\cite{Miroshnichenko2012, HopkinsPoddubny2013}.
The hybridization of electric dipoles is not sufficient for considering all-dielectric oligomers because high-index dielectric nanoparticles have both electric and magnetic dipolar responses~\cite{Evlyukhin2010, Kuznetsov2012, Evlyukhin2012NL}.  
To investigate the eigenmodes of such systems, we can again employ our simplified approach to hybridization, but instead use the coupled electric and magnetic dipole approximation~\cite{Mulholland1994}, which will account for both electric and  magnetic responses of individual nanoparticles.
From geometric symmetry principles, we know that this approach will result in two distinct sets of eigenmodes for the trimer: one which will transform under the trimer's symmetry operations according to the $\mathrm{E'}$  irreducible representation and the other which will transform according the $\mathrm{E''}$ irreducible representation (see Table~\ref{tab:D3h}).
The intuitive distinction between the two sets of eigenmodes is that they will be excited by either the electric or magnetic field in a normally-incident plane wave, for $\mathrm{E'}$ and $\mathrm{E''}$, respectively.
The different irreducible representations also means that the trimer's symmetry prevents $\mathrm{E'}$ eigenmodes coupling to the $\mathrm{E''}$ eigenmodes and {\it vice versa}.
If we could neglect the bianisotropic coupling between electric to magnetic dipoles, the $\mathrm{E'}$ eigenmodes would be those that we derived for plasmonic trimers ({\it i.e.} $|v_2\rangle$ and $|v_3 \rangle$ in Eq.~\ref{eq:plasmonic eigenmodes}) and the $\mathrm{E''}$ eigenmodes would be analogous to the same eigenmodes, but constructed from magnetic dipoles rather than electric dipoles. 
However, by using the out-of-plane ($z$) direction, the $z$-oriented magnetic dipoles can be arranged into basis vectors that transform according to the $\mathrm{E'}$ irreducible representation and the  $z$-oriented electric dipoles can be arranged into basis vectors that transform according to the $\mathrm{E''}$ irreducible representation, as shown in Fig.~\ref{fig:new_eigenmodes}b.  
Hence, bianisotropic coupling between electric and magnetic dipoles will allow the electric dipole $\mathrm{E'}$ basis vectors to couple with the magnetic dipole $\mathrm{E'}$ basis vectors, and {\it vice versa} for the $\mathrm{E''}$ basis vectors.
From this point on, we shall consider only the  $\mathrm{E'}$  eigenmodes  because they are the eigenmodes excited by electric field and can therefore be more naturally related to the eigenmodes we derived for plasmonic trimers.  
The procedure for finding the $\mathrm{E''}$ eigenmodes is almost identical upon interchanging the electric and magnetic dipole polarizabilities.
We begin by extending the hybridization diagram of Fig.~\ref{fig:electric_eigenmodes}d  to take into account both the new space of $z$-oriented $\mathrm{E'}$ responses, and the bianisotropic coupling channels.
In Fig.~\ref{fig:new_eigenmodes}a-c, we show the complete set of basis vectors that hybridize to form the  $\mathrm{E'}$  and  $\mathrm{E''}$ eigenmodes of an all-dielectric trimer.
These basis vectors can be separated according to their even or odd response under the dimer's reflection symmetry operation, because there are no coupling channels between the two resulting sets of basis vectors.  
One of these two sets is then sufficient to find instances of each doubly-degenerate eigenmode ({\it i.e.} being the instances that are even/odd under the dimer's reflection symmetry operation).
The even basis vectors offer a convenient opportunity to reduce the number of coupling channels because the trimer basis vector is an (orthogonal) eigenmode for the dimer.
The new, $E$ and $F$, coupling coefficients, shown in Fig.~\ref{fig:new_eigenmodes}d, can then be calculated by evaluating the magnetic and electric field radiated by electric and magnetic dipoles.  See Appendix~\ref{appendixB} for details.
\begin{subequations}
\begin{align}
& \begin{array}{cc}
 E_1 \;
=\;\alpha_{\scriptscriptstyle H}\frac{1}{\sqrt{\epsilon_0 \mu_0 }}\;\frac{ e^{i k R }}{4 \sqrt{2}\,  \pi  R}\left(k^2 + \frac{ i k}{ R}  \right) \hfill \\
E_2 \;
=\; -\alpha_{\scriptscriptstyle E}\sqrt{\epsilon_0 \mu_0 }\;\frac{e^{i k R }}{4   \sqrt{2}\,\pi  R}\left(k^2 + \frac{i k}{ R} \right)  
\end{array} \\
&\;\;\, F \;=\; \sqrt{2}\, E 
\end{align}
\label{eq:coupling' extra}
\end{subequations}
We can then define an eigenmode of the dielectric trimer, $\left | w \right \rangle$, as a linear combination of single particle and dimer eigenmodes.  
\begin{align}
\left | w_i \right \rangle = a_i'\left | \mathrm{p_y} \right \rangle+ b_i' \left | \mathrm{B_{2u}} \right \rangle + c_i' \left |  \mathrm{A_{g}} \right \rangle + d_i' \left | \mathrm{E'_z} \right \rangle 
\end{align}
Notably, the expression for the $\mathrm{A'_1}$ hybridized eigenmode of the trimer, which has radially-oriented electric dipole moments, is the same as $\left | v_4 \right \rangle$ in Eq.~\ref{eq:v4}.  
This is because the eigenmode is unable to couple bianisotropically into any magnetic dipoles and is, therefore, unchanged from the previous analysis provided for plasmonic trimers.  
The remaining three-dimensional eigenspace can therefore be spanned by the two  $\mathrm{E'}$  eigenmodes in  Eq.~\ref{eq:result} ($\left | v_{2y} \right \rangle$ and $\left | v_{3y} \right \rangle$) and the  $|\mathrm{E'_z}\rangle$ basis vector.
As such, all the remaining eigenmodes can be written as a linear combination of these three basis vectors, and, therefore, all remaining eigenmodes must transform according to the $\mathrm{E'}$ irreducible representation.  
Additionally, since the $\mathrm{E'}$ eigenmodes transform according to a different irreducible representation to that of $\left | v_4 \right \rangle$; we know they must be orthogonal to $\left | v_4 \right \rangle$, and we can thereby write their general form as:
\begin{align}
\left| {w_{i}} \right \rangle:\quad& \left \{ 
\begin{array}{ll}
a'_i &= a' \\
b'_i &= \sqrt{2}a' - \sqrt{3}c' \\
c'_i &= c'\\
d'_i &= d'
\end{array}\right.
\end{align}
If the bianisotropic coupling channels are negligible ($E,\,F \rightarrow 0$), the three basis vectors, $\left | v_{2y} \right \rangle$, $\left | v_{3y} \right \rangle$ and $|\mathrm{E'_z}\rangle$, are the $\mathrm{E'}$ eigenmodes of the trimer.
However, outside of this limit, we can use the dipole model to set up a rank 3 matrix equation to find the eigenmode solutions for $a'$, $c'$ and $d'$ (refer to Appendix B).  
These eigenmodes can be calculated analytically, but the result is cumbersome and does not provide any additional intuitive understanding.  
Therefore, it is sufficient to use a numerical approach, and this is what we do in the coming analysis.    
It is, however, worth acknowledging a conclusion here: because we have three linearly independent $\mathrm{E'}$ basis vectors, a dielectric trimer will have {\it three} $\mathrm{E'}$  eigenmodes, which one more than the plasmonic trimer.
Moreover, using the analogous argument, a dielectric trimer additionally has another three  $\mathrm{E''}$ eigenmodes that can be excited by the incident magnetic field.
Therefore, the total number of, doubly-degenerate, eigenmodes that can be excited in a dielectric trimer with a normally-incident plane wave is six, which are organized into two sets of three interfering eigenmodes. 
For comparison, a plasmonic heptamer (a central nanoparticle surrounded by a ring of six nanoparticles) has a single set of three interfering eigenmodes\cite{HopkinsPoddubny2013}.
The dielectric trimer geometry therefore has {\it double} the propensity for Fano resonances as the plasmonic heptamer geometries.  
In this regard, the theoretical predictions for the existence of Fano resonances in all-dielectric trimers were originally made for silicon spherical nanoparticles~\cite{Miroshnichenko2012, HopkinsPoddubny2013}, which exhibited Fano resonances in the optical frequency range.
However, it is possible to create a macroscopic analogue of silicon nanospheres in the microwave range, specifically: using MgO-$\mathrm{TiO_2}$ ceramic spheres characterized by dielectric constant of 16 and dielectric loss factor of $(1.12  -  1.17)\times10^{-4}$ (measured at $9-12\,\mathrm{GHz}$). 
These ceramic spheres in the microwave range therefore have very similar properties to silicon spheres in the optical range.  
As such, we are able to create a macroscopic analogue of a silicon nanosphere trimer.
This allows us to investigate the scattering properties of a single trimer with much less noise than would be possible for a single silicon nanosphere trimer. 
In Fig.~\ref{fig:experiment}a, we show an experimental setup. 
The trimer consists of three MgO-$\mathrm{TiO_2}$ spheres with $15\,\mathrm{mm}$ diameter and 20 mm distance between the centers of the spheres. 
To fasten together the  MgO-$\mathrm{TiO_2}$ particles for the experiment, we used a custom holder made of a styrofoam material with dielectric permittivity of 1 (in the microwave frequency range). 
To approximate plane wave excitation and receive the signal scattered to the forward direction, we employ a pair of, identical, rectangular linearly-polarized wideband horn antennas (TRIM, $\mathrm{1 - 18\,GHz}$), connected to the coaxial ports of a vector network analyzer (Agilent E8362C).
The trimer is located in the far-field of both antennas; the distance from the trimer to both the receiving and the transmitting antennas is approximately $\mathrm{1.5\,m}$. 
The total extinction can then be extracted from the measured complex magnitude of the forward scattered signal by means of the optical theorem~\cite{Larsson}.
Associated simulations of the experiment were also performed by using the Time Domain solver of CST Microwave Studio when assuming plane wave excitation on the trimer in free space. 
The experimentally measured, and numerically simulated,  extinction spectra are shown in Fig.~\ref{fig:experiment}b and \ref{fig:experiment}c.
The extinction spectrum of an isolated MgO-$\mathrm{TiO_2}$ sphere was also measured and simulated.
In Fig.~\ref{fig:experiment}b, we can see a pronounced Fano resonance at $4.8\,\mathrm{GHz}$ that is associated with the suppression of extinction.  By varying the orientation of the trimer, it exhibits a polarization independent response. 
The Fano resonance must subsequently be from interference between doubly-degenerate eigenmodes, and we can, therefore, consider the hybridized eigenmodes derived according to our previous analysis.
For this hybridization, we define electric and magnetic dipole polarizabilities from the $a_1$ and $b_1$ scattering coefficients of Mie theory~\cite{Mie1908}.  
The MgO-$\mathrm{TiO_2}$ permittivity was assumed be dispersionless with a dielectric constant of 16 and dielectric loss factor of $(1.12  -  1.17)\times10^{-4}$.
By calculating the hybridized eigenmodes we get the extinction spectra shown in Fig.~\ref{fig:theory}a, which accurately reproduces the experiment's extinction spectrum and Fano resonance.
The discrepancy between the hybridzation theory and experiment at high frequencies is due to the prescence of a known~\cite{Kuznetsov2012, Evlyukhin2012NL} magnetic quadrupole response in the individual spheres, which we simply do not take into account in our hybridization.  
Regarding, however, the Fano resonance at $4.8\,\mathrm{GHz}$, we are able to decompose the simulated extinction spectrum into components coming from each isolated eigenmode as seen in Fig.~\ref{fig:theory}.  
The extinction depicted for each eigenmode is `isolated' in the sense that we are neglecting the extinction that can be attributed to the interference between eigenmodes\cite{HopkinsPoddubny2013}.  
The decomposition clearly shows that there are multiple eigenmodes that interfere destructively to form the main Fano resonance at $4.8\,\mathrm{GHz}$.  
Moreover, modal interference is occurring at very similar frequencies in both the $\mathrm{E'}$ and $\mathrm{E''}$  response;  the Fano feature is coming from interference between $ | \mathrm{E'_{(2)}}  \rangle$ and $ |\mathrm{E'_{(3)}}  \rangle$, {\it and} between $ | \mathrm{E''_{(1)}}  \rangle$ and $ |\mathrm{E''_{(2)}}  \rangle$. 
It is worthwhile emphasizing that this is a simultaneous overlap of two optical Fano resonances, which are symmetrically-exclusive by nature of their distinct irreducible representations.
The likely reason for this situation is that $4.8\,\mathrm{GHz}$ is also the frequency of an individual sphere's magnetic dipole resonance, and at least one interfering eigenmode in both the $\mathrm{E'}$ and $\mathrm{E''}$  responses is dominated by magnetic dipoles.  
In other words, we have eigenmodes in both irreducible representations that are dependent on a single parameter; the magnetic dipole polarizability of a single MgO-$\mathrm{TiO_2}$ sphere.    
However, more generally, the correlation between a resonance of the single particle and modal interference is a recurring feature of the experiment.  
The single particle's electric resonance at $6.5\,\mathrm{GHz}$ coincides with destructive interference in the $\mathrm{E'}$ response, between the $ |\mathrm{E'_{(1)}}  \rangle$ and $ |\mathrm{E'_{(3)}}  \rangle$ eigenmodes. 
In Fig.~\ref{fig:experiment}b and \ref{fig:experiment}c, the single sphere's magnetic quadrupole resonance at $\mathrm{7\,GHz}$ is also associated with a Fano resonance feature.
It is interesting that the eigenmodes in our $\mathrm{E''}$ response exhibit, anomalistic, resonant behavior in the vicinity of the (neglected) magnetic quadrupole resonance.  
Indeed it is very likely that this frequency range should have significant mutlipolar coupling channels~\cite{Evlyukhin2011} that have been omitted in our hybridization. 
However, for the purposes of the work here, the key result is the largest Fano resonance at $4.8\,\mathrm{GHz}$, which is fully described by our hybridization theory.
This Fano resonance is a realization of the predicted propensity that all-dielectric trimers have towards Fano resonances.

\section{Conclusions}

We have presented an explicit study on the hybridization of optical responses in both plasmonic and all-dielectric trimers.
We presented a simplified hybridization model to allow us to derive the eigenmodes of these structures analytically and observed the formation of modal interference in eigenmodes excited by normally-incident excitation. 
In this regard, plasmonic trimers were shown to exhibit nonorthogonality from retarded coupling channels, and dielectric trimers could further utilize bianisotropic coupling channels. 
A key prediction of our hybridization theory was then demonstrated experimentally: an all-dielectric trimer was shown to exhibit sharp, polarization-independent Fano resonances.
The measurements were in good agreement with our simplified hybridization model, to therefore validate our approach.   
Our conclusions subsequently demonstrate that the use of full plasmonic hybridization is not necessary to deduce the dominant eigenmodes of multiparticle geometries.  
Yet, importantly, our analysis has demonstrated that hybridized eigenmodes can only be related to the set of intercoupled basis vectors in a simple form if the basis vectors are, in isolation, eigenmodes of their associated subsystem.  
Indeed, this raises a more general point of contention against many empirical applications of hybridization concepts and establishes why a more considered utilization of hybridization concepts in terms of subsystem eigenmodes is necessary.

\section{Acknowledgements}
This work was supported by the Australian Research Council. The measurements were supported by the Government of the Russian Federation (grant 074-U01), the Ministry of Education and Science of the Russian Federation, Russian Foundation for Basic Research, Dynasty Foundation (Russia).  BH acknowledges a number of useful discussions with Guangyao Li, Daniel Leykam, Anton S. Desyatnikov and David A. Powell.

\appendix
\section{Plasmonic trimer} \label{appendixA}
The dipole model we use for a plasmonic trimer assumes only electric dipoles from the individual nanoparticles.
\begin{align}
\mathbf{p}_{i}  = \alpha_{\scriptscriptstyle E}\epsilon_{0}\mathbf{E_{0}}(\mathbf{r}_{i})&+\alpha_{\scriptscriptstyle E}k^{2}
\sum \limits_{j\neq i} \hat{G}_{0}(\mathbf{r}_{i},\mathbf{r}_{j}) \cdot \mathbf{p}_{j}
\end{align}
Here, $\alpha_{\scriptscriptstyle E}$ is the electric dipole polarizability of a particle and $\hat{G}_{0}$ is the free space dyadic Green's function, which acts on dipole moments as:
\begin{align}
\hat{G}_0(\mathbf{r'},\mathbf{r})\cdot \mathbf{p} 
&= \frac{e^{i k R }}{4 \pi   R}\left[ \left(1 + \frac{i}{kR}- \frac{1}{k^2 R^2}\right) \mathbf{p} - \left( 1 + \frac{3i}{kR}- \frac{3}{k^2R^2}\right) (\mathbf{n}\cdot\mathbf{p})\mathbf{n}  \right ] \nonumber
\end{align}
where $k$ is the wavenumber, $\mathbf{n}$ is the unit vector pointing from $\mathbf{r}$ to $\mathbf{r'}$ and $R = |\mathbf{r}-\mathbf{r'}|$.
The associated eigenmode equation for a plasmonic trimer in this dipole model can then be written as:
\begin{align}
\mathbf{v}_i = \alpha_{\scriptscriptstyle E} \epsilon_0 \lambda \mathbf{v}_i  +  \sum \limits_{j\neq i} \alpha_{\scriptscriptstyle E}k^2 \hat{G}_0(\mathbf{r}_i,\mathbf{r}_j)\cdot \mathbf{v}_j
\label{eq:eig equation app}
\end{align}
If we define our eigenmode using the dimer and single particle eigenmodes as basis vectors, we can write any associated eigenmode as:
\begin{align}
\left| {v_i} \right \rangle= a_i \left|\mathrm{p_x}\right \rangle + b_i\left|  \mathrm{B_{3u}}\right \rangle  + c_i \left|\mathrm{B_{1g}}\right \rangle +  d_i \left|\mathrm{p_y}\right \rangle + e_i\left|  \mathrm{B_{2u}}\right \rangle  + f_i \left|\mathrm{A_{g}}\right \rangle 
\end{align}
The six basis vectors are the dimer and single particle eigenmodes we define in the main text.
The  eigenmode equation, Eq.~\ref{eq:eig equation app}, can then be written as a $6\times6$ matrix equation in terms of these basis vectors and the associated coupling channels.  
\begin{align}
\left( \begin{array}{cccccc}
1  & -A & -B & 0 & 0 & 0\\
-A & \alpha_{\scriptscriptstyle E} \epsilon_0 \lambda_{\mathrm{B_{3u}}} &0&0&0&0\\
-B & 0 & \alpha_{\scriptscriptstyle E} \epsilon_0 \lambda_{\mathrm{B_{1g}}} &0&0&0\\
0&0&0&   1 & -C  & -D\\
0&0&0&  -C &  \alpha_{\scriptscriptstyle E} \epsilon_0 \lambda_{\mathrm{B_{2u}}}  & 0 \\
0&0&0&   -D & 0 & \alpha_{\scriptscriptstyle E} \epsilon_0 \lambda_{\mathrm{A_{g}}} 
\end{array}\right) 
\left( \begin{array}{c}
a_i\\b_i\\c_i\\d_i\\e_i\\f_i
\end{array}\right)  =
\alpha_E \epsilon_0 \lambda_i 
\left( \begin{array}{c}
a_i\\b_i\\c_i\\d_i\\e_i\\f_i
\end{array}\right)
\label{eq:Pmatrix}
\end{align}
where the coupling channels, $A$-$D$, are defined in Eq.~\ref{eq:coupling} of the main text and $\lambda_{\mathrm{B_{3u}}}$, $\lambda_{\mathrm{B_{1g}}}$, $\lambda_{\mathrm{B_{2u}}}$ and $\lambda_{\mathrm{A_{g}}}$ are the eigenvalues of the corresponding eigenmodes in the isolated dimer.  
\begin{align}
 \lambda_{\mathrm{B_{3u}}} &= (\alpha_{\scriptscriptstyle E} \epsilon_0)^{-1} + \frac{ e^{i k R }}{2 \pi  \epsilon_0 R}\left( \frac{ik}{R}-\frac{1}{R^2}\right)
 \\
  \lambda_{\mathrm{B_{1g}}} &= (\alpha_{\scriptscriptstyle E} \epsilon_0)^{-1}  + \frac{e^{i k R }}{4 \pi \epsilon_0 R} \left( k^{2} + \frac{ik}{R}- \frac{1}{R^2}\right)
\\ 
\lambda_{\mathrm{B_{2u}}} &= ( \alpha_{\scriptscriptstyle E} \epsilon_0)^{-1} - \frac{e^{i k R }}{4 \pi \epsilon_0 R} \left( k^{2} + \frac{ik}{R}- \frac{1}{R^2}\right) 
  \\
 \lambda_{\mathrm{A_{g}}} &=  ( \alpha_{\scriptscriptstyle E} \epsilon_0)^{-1} - \frac{ e^{i k R }}{2 \pi \epsilon_0  R} \left( \frac{ik}{R}- \frac{1}{R^2}\right)
\end{align}
These eigenvalues account for self-interaction of the dimer eigenmodes in the above matrix equation.  
The six solutions of Eq.~\ref{eq:Pmatrix} provide the eigenmodes and eigenvalues for the plasmonic trimer in the main text.  
However, in regard to the nonorthogonality of $| v_2 \rangle$ and $| v_3 \rangle$, it is worth acknowledging that this nonorthogonality requires retarded coupling between the particles.
Moreover, if refer to the matrix in Eq.~\ref{eq:Pmatrix} and the definitions of coupling channels in Eq.~\ref{eq:coupling}, we note that by neglecting retardation in the coupling between dipoles, we make the phase acquired through all coupling channels simply equal to that provided by the dipole polarizability.  
In other words, $\frac{A}{\alpha_E}$, $\frac{B}{\alpha_E}$, $\frac{C}{\alpha_E}$ and $\frac{D}{\alpha_E}$ all become real numbers. 
This then makes the matrix in Eq.~\ref{eq:Pmatrix} proportional to a real symmetric matrix, and hence Hermitian, which then guarantees it has orthogonal eigenvectors.  
As we already have orthonormal basis vectors, this then makes all eigenmodes orthogonal.  
As such this shows that the retardation of coupling between particles is necessary for nonorthogonal eigenmodes.

\section{Dielectric trimer}  \label{appendixB}

For modeling the dielectric trimer, we use the coupled electric and magnetic dipole approximation, which is described by the following two (coupled) equations:
\begin{subequations}\label{eq:Dmodel}
\begin{align}
\mathbf{p}_{i}  = \alpha_{E}\epsilon_{0}\mathbf{E_{0}}(\mathbf{r}_{i})&
+\alpha_{E}k^{2}
  \left(\underset{j\neq i}{{\sum}}\hat{G}_{0}(\mathbf{r}_{i},\mathbf{r}_{j})\cdot  \mathbf{p}_{j}
-\frac{1}{c_0}\,\nabla\times\hat{G}_{0}(\mathbf{r}_{i},\mathbf{r}_{j}) \cdot \mathbf{m}_{j}\right) \\
 \mathbf{m}_{i} =  \alpha_{H}\mathbf{H_{0}}(\mathbf{r}_{i})&
 +\alpha_{H}k^{2}  \left(\underset{j\neq i}{{\sum}}\hat{G}_{0}(\mathbf{r}_{i},\mathbf{r}_{j}) \cdot \mathbf{m}_{j}
+c_0\,\nabla\times\hat{G}_{0}(\mathbf{r}_{i},\mathbf{r}_{j}) \cdot \mathbf{p}_{j}\right)
\end{align}
\end{subequations}
where $\mathbf{p}_i$ ($\mathbf{m}_i$) is the electric (magnetic) dipole moment of the $i^{\mathrm{th}}$ particle, $\hat{G}_{0}(\mathbf{r}_{i},\mathbf{r}_{j}) $ is the free space dyadic Greens function between the $i^{\mathrm{th}}$ and $j^{\mathrm{th}}$ dipole, $\alpha_{E}$ ($\alpha_{H}$) is the electric (magnetic) polarizability of a particle, $c_0$ is the speed of light and $k$ is the free-space wavenumber. 
The extra bianisotropic coupling terms are given according to:
\begin{align}
\nabla\times\hat{G}_{0}(\mathbf{r}',\mathbf{r}) \cdot \mathbf{p} 
&=\frac{ e^{i k R }}{4 \pi  R}\left(1 + \frac{i}{k R}  \right) \mathbf{n}\times\mathbf{p}
\end{align}
An eigenmode of the dielectric trimer, having electric dipoles $\mathbf{v}$ and magnetic dipoles $\mathbf{u}$, will therefore satisfy the coupled electric and magnetic dipole model (Eq.~\ref{eq:Dmodel}) as:
\begin{subequations}\label{eq:Deigen}
\begin{align}
\mathbf{v}_{i}  = \alpha_{E}\epsilon_{0} \lambda \mathbf{v}_i&
+\alpha_{E}k^{2}
  \left(\underset{j\neq i}{{\sum}}\hat{G}_{0}(\mathbf{r}_{i},\mathbf{r}_{j})\cdot  \mathbf{v}_{j}
-\frac{1}{c_0}\,\nabla\times\hat{G}_{0}(\mathbf{r}_{i},\mathbf{r}_{j}) \cdot \mathbf{u}_{j}\right) \\
 \mathbf{u}_{i} =  \alpha_{H}\lambda \mathbf{u}_i&
 +\alpha_{H}k^{2}  \left(\underset{j\neq i}{{\sum}}\hat{G}_{0}(\mathbf{r}_{i},\mathbf{r}_{j}) \cdot \mathbf{u}_{j}
+c_0\,\nabla\times\hat{G}_{0}(\mathbf{r}_{i},\mathbf{r}_{j}) \cdot \mathbf{v}_{j}\right)
\end{align}
\end{subequations}
We can then use the dimer and single particle basis vectors from Fig.~\ref{fig:new_eigenmodes} in the main text to write an expression that will describe the eigenmodes of a dielectric trimer that are invariant under the dimer's reflection symmetry operation.
\begin{align}
\left | w_i \right \rangle = a_i'\left | \mathrm{p_y} \right \rangle+ b_i' \left | \mathrm{B_{2u}} \right \rangle + c_i' \left |  \mathrm{A_{g}} \right \rangle + d_i' \left | \mathrm{E'_z} \right \rangle 
\end{align}
Analogous to the case for plasmonic trimers in Appendix~\ref{appendixA}, we can rewrite Eq.~\ref{eq:Deigen} as a $4\times4$ matrix equation.
\begin{align}
\left( \begin{array}{cccc}
1  & -C & -D & -E_2 \\
-C & \alpha_{\scriptscriptstyle E} \epsilon_0 \lambda_{\mathrm{B_{2u}}} &0&-F_2\\
-D & 0 & \alpha_{\scriptscriptstyle E} \epsilon_0 \lambda_{\mathrm{A_{g}}} &0\\
-E_1 & - F_1  & 0 & \alpha_H \lambda_{\mathrm{E'_z}} 
\end{array}\right) 
\left( \begin{array}{c}
a'_i\\b'_i\\c'_i\\d'_i
\end{array}\right)  =
 \lambda_i \left( \begin{array}{c}
\alpha_{\scriptscriptstyle E}  \epsilon_0 a'_i\\ \alpha_{\scriptscriptstyle E}  \epsilon_0 b'_i\\ \alpha_{\scriptscriptstyle E}  \epsilon_0 c'_i\\ \alpha_H d'_i
\end{array}\right)
\label{eq:Dmatrix}
\end{align}
where the coupling channels, $C$-$F$, are defined in Eqs.~\ref{eq:coupling} and \ref{eq:coupling' extra} of the main text and the eigenvalue of the dimer eigenmode with $z$-oriented magnetic dipole moments is:
\begin{align}
\lambda_{\mathrm{E'_z}} &=  \alpha_{\scriptscriptstyle H}^{-1} + \frac{ e^{i k R }}{4 \pi  R} \left( k^{2} + \frac{ik}{R}- \frac{1}{R^2}\right) 
\end{align}
We can then reduce Eq.~\ref{eq:Dmatrix} to a $3\times3$ matrix equation when searching for only the doubly-degenerate eigenmodes given the eigenmode with radially-oriented dipole moments, $|v_4\rangle$, remains an eigenmode of Eq.~\ref{eq:Dmatrix}. 
This is done by considering the eigenspace that is orthogonal to $|v_4\rangle$.
In effect, this means we substitute $b'_i = \sqrt{2}a'_i - \sqrt{3}c'_i$ into Eq.~\ref{eq:Dmatrix} to get the reduced matrix equation:
\begin{align}
\left( \begin{array}{ccc}
1 - \sqrt{2}C   & \sqrt{3}C-D  & -E_2 \\
-D & \alpha_{\scriptscriptstyle E} \epsilon_0 \lambda_{\mathrm{A_{g}}} &0\\
 -E_1 - \sqrt{2}F_1 & \sqrt{3}F_1 & \alpha_H \lambda_{\mathrm{E'_z}} 
\end{array}\right) 
\left( \begin{array}{c}
a'_i\\c'_i\\d'_i
\end{array}\right)  =
 \lambda_i \left( \begin{array}{c}
\alpha_{\scriptscriptstyle E}  \epsilon_0 a'_i\\  \alpha_{\scriptscriptstyle E}  \epsilon_0 c'_i\\ \alpha_H d'_i
\end{array}\right)
\label{eq:Dmatrixreduced}
\end{align}
From this expression, it is straightforward to find numerical solutions to this matrix equation and obtain both the eigenmodes and eigenvalues.   
However, as an addendum, we can consider the relations between the  $a'$, $b'$ and $c'$ coefficients of eigenmodes analytically.  
This will allow us to illustrate the interdependences of the collective eigenmodes on the resonances of the individual consitituent subsystems. 
Moreover, we can use the second row of Eq.~\ref{eq:Dmatrixreduced} to first write a relationship between $a'$ and $c'$ coefficients:
\begin{align}
-D a'_i = (\lambda_i  -\lambda_{\mathrm{A_g}} )\alpha_E \epsilon_0 c'_i
\label{meow}
\end{align}
From this relation, and using $F_1 = \sqrt{2}E_1$, the third row of Eq.~\ref{eq:Dmatrixreduced} can also provide a relationship between $c'$ and $d'$ coefficients:
\begin{align}
(\lambda_i-\lambda_{\mathrm{E'_z}} )  D \alpha_H d'_i  = E_1 (3 (\lambda_i  -\lambda_{\mathrm{A_g}} )\alpha_E \epsilon_0  + D\sqrt{6})c'_i 
\label{woof}
\end{align}
These two relations are not sufficient to define eigenmodes analytically, because we need the first row of Eq.~\ref{eq:Dmatrixreduced} to solve for $\lambda_i$, which becomes quite nontrivial.  
However, for our purposes, Eq.~\ref{meow} and \ref{woof} are sufficient to define the profile of an eigenmode for its given eigenvalue.  
Indeed, these two relations nicely illustrate the how the hybridized eigenmodes become dependent on each basis vector's resonant properties and all the coupling channels. 
\bibliographystyle{apsrev4-1}
\bibliography{bibliography}

\end{document}